\documentclass[aps,pra,showpacs,twoside,twocolumn,longbibliography,10pt]{revtex4-1}
\usepackage[colorlinks=true, citecolor=red, urlcolor=blue ]{hyperref}
\usepackage{epsfig,newlfont,amssymb,amsfonts,amsmath,bm,subfigure,palatino,mathtools,amsthm,braket,times,soul,enumitem,color}
\usepackage[normalem]{ulem}
\newcommand{\stkout}[1]{\ifmmode\text{\sout{\ensuremath{#1}}}\else\sout{#1}\fi}
\usepackage[english]{babel}
\usepackage[utf8]{inputenc}
\usepackage{array}
\usepackage{xcolor}
\usepackage{graphics}

\newcommand{\ketbra}[2]{|#1\rangle \langle #2|}
\def\Tr{\text{Tr}}

\usepackage{amsthm}
\usepackage{verbatim}
\usepackage{bbm}
\usepackage{wrapfig}
\usepackage{makecell}
\usepackage{tikz}
\usetikzlibrary{arrows.meta, decorations.pathreplacing, calc, shapes.geometric}

\usepackage{orcidlink}

\usepackage{hyphenat}
    \usepackage{amsmath}

\newlength\figureheight 
\newlength\figurewidth 

\begin{document}

\title{Memory-assisted advantage for state transfer in disordered quantum many-body scar system}

\author{Paranjoy Chaki \orcidlink{0009-0000-5693-5516} }
\affiliation{Harish-Chandra Research Institute,  
Chhatnag Road, Jhunsi, Prayagraj 211 019, India\\
Homi Bhabha National Institute,  Training School Complex, Anushakti Nagar, Mumbai 400 094, India}

\author{Ujjwal Sen \orcidlink{0000-0002-0091-5847}}
\affiliation{Harish-Chandra Research Institute,  
Chhatnag Road, Jhunsi, Prayagraj 211 019, India\\
Homi Bhabha National Institute,  Training School Complex, Anushakti Nagar, Mumbai 400 094, India}
\begin{abstract}


We analyze how memory in disorder facilitates quantum communication in many-body scar systems. 
%
%
We consider three distinct types of disorder, viz., memoryful, and memoryless uniform and Gaussian, and compare their respective performances in facilitating quantum state transfer. Using the maximum transfer fidelity and fidelity area as figures of merit, we find that memoryful disorder yields a better performance than the memoryless disordered channels.
Furthermore, the maximum transfer fidelity exhibits an initial parabolic decay with disorder strength, followed by a linear decrease, for all the disorder models considered. We introduce 
a 
degree of scarness, and show that it is higher for memoryful disorder in comparison to  memoryless disorders, implying a role of scarness in the quantum state transfer protocol. We further perform a scaling analysis, revealing that memory effect in disorder is not only beneficial for short-distance but also long-distance quantum state transfer. 
Finally, we show that the state yielding the maximum transfer fidelity has larger inverse participation ratio for memoryful disorder in comparison to the other two disorders, highlighting the role of nonergodicity in enhancing state transfer.
\end{abstract}

\maketitle

\section{Introduction}
Quantum state transfer (QST) is a vital task in quantum information science that aims to reliably transmit an unknown quantum state between distant nodes of a quantum network. Efficient quantum state transfer protocols play a crucial role in quantum communication and scalable quantum technologies. To the best of our knowledge, quantum state transfer protocols based on the dynamics of many-body Hamiltonians were first proposed in Ref.~\cite{unm}. Subsequently, significant developments have been achieved in this direction, as discussed in Refs.~\cite{QST_1,QST_2,QST_3,QST_4,QST_5,QST_6,QST_7,QST_8}. In parallel, quantum state transfer has also been extensively investigated in non-Hermitian systems~\cite{arx_nh}, long-range interacting spin systems~\cite{lg}, photonic platforms~\cite{photon_1,Photon_2}, ultracold atomic systems~\cite{ultra_cold_atom}, superconducting cavities~\cite{Sup_cavity} etc.

In the above-mentioned references, the transfer of an unknown quantum state is achieved with a fidelity smaller than unity, and the transfer fidelity generally deteriorates with increasing system size. To circumvent this limitation, specially engineered qubit interactions have been proposed, enabling the transfer of an arbitrary quantum state over arbitrary distances up to a local phase factor that is independent of the system size. This phenomenon, known as "perfect state transfer," was first introduced in Ref.~\cite{PST} and has since attracted considerable attention due to its broad range of applications across diverse physical platforms~\cite{PST_1,PST_2,PST_2005,PsT_x1,Han_2025,PST_rev_2025,PST_2022}. One of the very important works in this direction is explored in one-dimensional chaotic many-body spin systems that contain scar eigenstates, which is given in Ref.~\cite{PST_scar}. Quantum many-body scars~\cite{scar_1,scar_2,scar_3,scar_4,scar_5,scar_Arka} are atypical eigenstates of chaotic many-body quantum systems that violate the eigenstate thermalization hypothesis (ETH), despite being surrounded by other thermal eigenstates, thereby giving rise to weak ergodicity breaking. Apart from the quantum state transfer, quantum many-body scars have found a wide range of applications in quantum metrology~\cite{Metro_1,Metro_2}, quantum batteries~\cite{Battery}, and quantum simulation~\cite{sim_1}. In Ref.~\cite{PST_scar}, the authors show that although chaotic systems satisfying the eigenstate thermalization hypothesis (ETH) exhibit poor quantum state transfer performance and fail to surpass the classical fidelity threshold, the introduction of quantum many-body scar states enables perfect state transfer.

In this work, we consider a more realistic scenario by introducing disorder into a one-dimensional many-body scar system. Although disorder effects in quantum state transfer protocols have been extensively studied~\cite{Dis_01,Dis_1,dis_2,Dis_4,Dis_5,Dis_6}, the role of disorder in scarred many-body Hamiltonians in the context of quantum state transfer remains unexplored. To this end, we consider three different types of disorder: two of them are memoryless disorders, i.e., uniform and Gaussian disorders, and a third memoryful disorder, which is a time-dependent Gaussian disorder. Such a distribution arises in the continuum limit of the elephant quantum walk~\cite{elephant}. \begin{figure}
		\centering
	\includegraphics[scale=0.12]{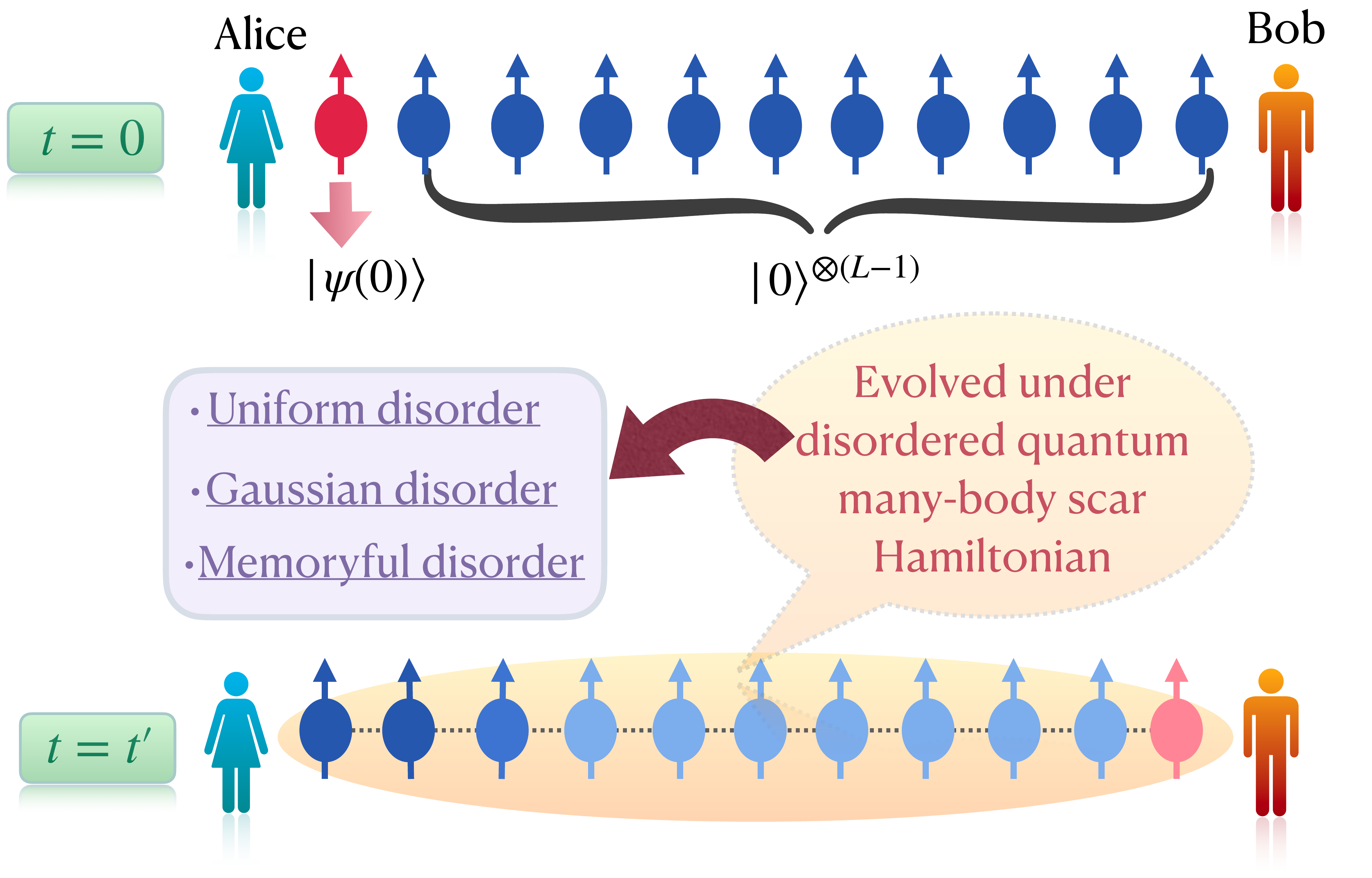}
		\caption{\textbf{Schematic representation of quantum state transfer through a disordered many-body scar system.} In this schematic diagram, we present a pictorial illustration of quantum state transfer through  a disordered many-body scar system. Alice aims to transfer her state $\ket{\psi(0)}$ to Bob through the time evolution governed by the disordered many-body scar Hamiltonian. Here, the total system size is $L$, and the remaining $L-1$ spins in the chain are initially prepared in the state $\ket{0}^{\otimes (L-1)} $. 
}
		\label{Fig1}
	\end{figure} The advantages of memory in quantum information processing and quantum technology are investigated in several aspects~\cite{Qunt_revv}. In this work, we examine how memory effects in disorder can be exploited to improve quantum state transfer through the dynamics, governed by a disordered many-body scar Hamiltonian. To characterize the performance of the state transfer protocol, we employ two figures of merit: the maximum transfer fidelity, defined as the maximum value of the fidelity over the time domain, and the fidelity area, which quantifies the area enclosed by the dynamical fidelity curve above the classical fidelity. We show that, in the presence of disorder, the state transfer protocol no longer remains perfect, i.e., an unknown quantum state can no longer be transferred to an arbitrary distance with unit fidelity. Additionally, for all three types of disorder, we observe the decay of maximum transfer probability falls parabolically up to a certain disorder strength, and then it decreases in a linear fashion. Therefore, it reveals a parabolic to linear transition in the decay of maximum fidelity. with disorder strength. At the same time, we notice among the three disorders, memoryful disorder performs better in comparison to the other two memoryless disorders by posing higher maximum transfer fidelity over the considered range of disorder strength than the other two types of disorder, showing that the presence of memory in a disorder provides clear benefits in a quantum state transfer protocol. Between uniform and Gaussian disorder uniform disorder shows an advantage in comparison to Gaussian disorder. At the same time, we get the similar feature in terms of fidelity area, i.e., the fidelity area of memory disorders remains higher in comparison to other disorders for all considered disorder strengths, and between uniform and Gaussian disorders, the value of area fidelity remains high, corresponding to uniform disorders. 

Next, to investigate how the scarness of the system varies with disorder strength, we introduce a measure of scarness based on the periodic revival of local observables~\cite{scar_1} in the time dynamics and term it as the degree of scarness. For all three types of disorder, the degree of scarness initially decreases non-linearly with increasing disorder strength up to a strength, beyond which it decays linearly. Interestingly, the comparative behavior among the three disorders closely resembles that observed for the maximum fidelity. In particular, the degree of scarness remains highest in the presence of memoryful disorder, followed by memoryless uniform disorder and then memoryless Gaussian disorder. This observation suggests that a higher degree of scarness is associated with enhanced state-transfer performance. After that, to catch the decay behavior of maximum fidelity and fidelity area with system sizes, we perform a finite-size scaling analysis of the maximum fidelity and fidelity area and observe $\log$-$\log$ scaling behavior for both quantities across all considered disorder types. We note that the fall of both the quantities with system size is slower for memoryful disorder in comparison to memoryless uniform and Gaussian disorders. Finally, we consider the inverse participation ratio (IPR) to gain deeper insight into the connection between nonergodicity and quantum state transfer in our model. The IPR is a widely used for characterizing localization-delocalization  behavior in quantum many-body systems~\cite{IPR_11,IPR_122,IPR_1,IPR_2,IPR_L,IPR_LL,loc_ap} and also as a relevant quantity in state transfer protocols~\cite{Dis_4,Dis_6}. Here,we calculate the inverse participation ratio of the state that provides the maximum fidelity in time dynamics and the average inverse participation ratio over the inverse participation ratios of all the states that provide the transfer fidelity more than classical fidelity. For all three disorders, both $\text{IPR}_M$ and $\text{IPR}_A$ decrease with the disorder strength. On the other hand, the memoryful disorder yields higher values of both the quantities compared to the uniform disorder, while the uniform disorder performs better than the Gaussian disorder. This indicates that ergodic behavior becomes increasingly pronounced with increasing disorder strength and leading to a deterioration of the state transfer performance. 

The paper is organized as follows. Sec.~\ref{II} deals with the basic setup of our work and is divided into two subsections. Particularly, in Sec.~\ref{IIA}, we briefly introduce the quantum state transfer protocol, and in Sec.~\ref{IIB}, we present the disordered many-body scar Hamiltonian used in this work and discuss the three types of disorder considered in our study. In Sec.~\ref{III}, we demonstrate the beneficial role of memoryful disorder in the state transfer protocol. Subsequently, in Sec.~\ref{IV}, we introduce a measure of scarness and analyze its behavior with increasing disorder strength for all three types of disorder. In Sec.~\ref{SV}, we study the variation of the maximum fidelity and fidelity area with system size and present the corresponding finite-size scaling analysis of these quantities. In Sec.~\ref{VI}, we investigate the behavior of the inverse participation ratio (IPR) with disorder strength and discuss its influence on the state transfer protocol. Finally, we conclude our work in Sec.~\ref{VII}.

\section{setting the stage}\label{II}

In this section we discuss the prerequisites of quantum state transfer in a quantum scar system and our setup where we impose disorder in the one-dimensional many-body scar Hamiltonian.

\subsection{Quantum state transfer}\label{IIA}
Various studies have already been carried out in the context of QST. Investigating this task in many-body scar systems~\cite{PST_scar} has recently attracted significant attention due to their experimental realization. The Hamiltonian describing the 1D spin-$\frac{1}{2}$ many-body scar system employed in the quantum state transfer protocol is given by

\begin{eqnarray}
H
&=& \frac{h_z}{2} \sum_{n=1}^{L} \sigma^z_n 
+ \frac{1}{2} \sum_{n=1}^{L-1} \lambda_n 
\left( \sigma^x_n \sigma^x_{n+1} + \sigma^y_n \sigma^y_{n+1}  \right)\\ \nonumber
&+&\sum_{n=1}^{L-1} \hat{P}_n \hat{h}_n \hat{P}_n. 
\end{eqnarray}
Here, $L$ denotes the total number of spins in the chain. The operators $\sigma_x$, $\sigma_y$, and $\sigma_z$ represent the Pauli matrices. The site-dependent nearest-neighbor coupling strength is given by $\lambda_n=\sqrt{n(L-n)}$, where $\lambda$ is a constant parameter. The parameter $h_z$ denotes the strength of the external field applied along the $z$ direction.

At the same time, $h_n$ represents the three-body interaction term, which is given by {$h_n=\sigma_x\otimes\sigma_x\otimes\sigma_x$}, and $\hat{P}_n$ is a projector, defined as $\hat{P}_n=\mathbbm{I}^{\otimes3}_2-\ketbra{000}{000}-\ketbra{001}{001}-\ketbra{010}{010}-\ketbra{100}{100}$. The term $\sum_{n=1}^{L-1} \hat{P}_n \hat{h}_n \hat{P}_n$ introduces a kinetic constraint in the system such that $\hat{P}_n \hat{h}_n \hat{P}_n\ket{S}=0$, where $\ket{S}\in\{\ketbra{000}{000},\ketbra{100}{100},\ketbra{010}{010}, \ketbra{001}{001}\}$. As a result, this kinetic constraint effectively introduces the scar eigenstates into the system. 

Let us consider that the initial state of the entire 1D spin chain is given by $\ket{\psi}_{in}=\ket{\psi}\otimes\ket{0}^{\otimes L}$. If the system is allowed to evolve under the Hamiltonian $H$, then at time $\tau_m=\frac{\pi(m-1/2)}{2\lambda}$ the state becomes
\begin{equation}
   \ket{\psi(\tau_m)}=\exp(-iH\tau_m)\ket{\psi}_{in}=\ket{0}^{\otimes(L-1)}\otimes \hat{U}({\tau_m})\ket{\psi}\nonumber.
\end{equation}

Here, at time $\tau_m$ we can transfer any arbitrary state $\ket{\psi}$ from site $1$ to site $L$ with a local phase, which is independent of the initial state $\ket{\psi}$. The unitary operator $\hat{U}(t)$ corresponding to the local phase at any time $t$ is given by 
\begin{equation}
    \hat{U}(t)
=
\exp\left\{
i \left[
\frac{\lambda (L-1)+h_z}{2}\,\hat{Z}
+
h_z(L-1)\,\hat{I}_2
\right]t
\right\}\nonumber.
\end{equation}
Therefore, to receive the exact state at the site $L$, we apply the local unitary $\hat{U}^\dagger(t)$ on $L$'th site and the resulting transfer fidelity $
F(t) = \langle \psi \mid \hat{U}^\dagger(t)\, \hat{\rho}_N(t)\, \hat{U}(t) \mid \psi \rangle
$ fidelity become $1$ after each period $\Delta \tau=\tau_{m+1}-\tau_{m}=\frac{\pi}{2\lambda}$ starting from the initial configuration.

\subsection{Disordered many-body scar system}\label{IIB}

In this work, we investigate the effect of disorder on the QST protocol, where the presence of the disorder is considered specifically in the projector $P_n$, which is given by
\begin{eqnarray}\nonumber\label{2eq}
\hat{P}^{\delta}_n&=&\mathbbm{I}^{\otimes3}_2-(1+\delta_n)(\ketbra{000}{000}-\ketbra{001}{001}-\ketbra{010}{010}\\
&-&\ketbra{100}{100}).
\end{eqnarray}
Here, the parameter $\delta_n$ is a site-dependent quantity, sampled from a distribution of mean value $1$ and a given standard deviation (disorder strength) $\sigma$.  Therefore the Hamiltonian in presence of disorder is given by
\begin{eqnarray}\label{Eq.3}
H_D
&=& \frac{h_z}{2} \sum_{n=1}^{L} \sigma^z_n 
+ \frac{1}{2} \sum_{n=1}^{L-1} \lambda_n 
\left( \sigma^x_n \sigma^x_{n+1} + \sigma^y_n \sigma^y_{n+1}  \right)\\ \nonumber
&+&\sum_{n=1}^{L-1} \hat{P}^{\delta}_n \hat{h}_n \hat{P}^{\delta}_n. 
\end{eqnarray}
Let us consider that Alice wishes to send an unknown quantum state $\ket{\psi}=\ket{\psi(\theta,\phi)}=\cos(\frac{\theta}{2})\ket{0}+\exp(-i\phi)\sin(\frac{\theta}{2})\ket{1}$ from site $1$ to site $L$, where Bob is located. Here $\theta\in[0,\pi]$ and $\phi\in[0,2\pi]$. The remaining sites of the chain are initialized in the state $\ket{0}^{\otimes(L-1)}$. Therefore the total system is in an initial state, $\ket{\psi}_{in}=\ket{\psi(\theta,\phi)}\otimes\ket{0}^{\otimes L}$. After time $t$ the state of the whole system will be in a state $\ket{\psi(t)}=\exp(-iH_Dt)\ket{\psi}_{in}$ and the reduced density matrix $\rho_L(t)$ will be $\rho_L(t)=\Tr_{\bar{L}}[\ketbra{\psi(t)}{\psi(t)}]$. To nullify the effect of the local phase, we apply the unitary $\hat{U}^{\dag}(t)$ defined in i.e. $\rho'_L(t)=\hat{U}^{\dagger}(t)\rho'_L(t) \hat{U}(t)$.   Here we consider the average transfer fidelity, $\bar{F}$, where the average is taken over all the initial state $\ket{\psi(\theta,\phi)}$,  is given by

\[
\begin{aligned}
\bar{F} &= \frac{1}{4\pi} \int_{0}^{\pi} \int_{0}^{2\pi}
     \langle \psi(\theta,\phi) | \rho'_L(t) | \psi(\theta,\phi) \rangle
     \sin\theta \, d\theta \, d\phi .
\end{aligned}
\]
The state transfer protocol is considered effective when the average fidelity exceeds the classical threshold, i.e., $\bar{F}>F_c=2/3$. Here $F_c$ denotes the classical fidelity, which corresponds to the maximum fidelity achievable in the absence of a quantum channel during the state transfer. Therefore, the condition $\bar{F}>F_c$ signifies the quantum advantage. For the sake of simplicity, from now onward we refer to the average transfer fidelity simply as the fidelity. The pictorial description of state transfer protocol is given in Fig.~\ref{Fig1}.

\subsubsection{Disorder with memory effect}
\begin{figure}
		\centering
	\includegraphics[scale=0.38]{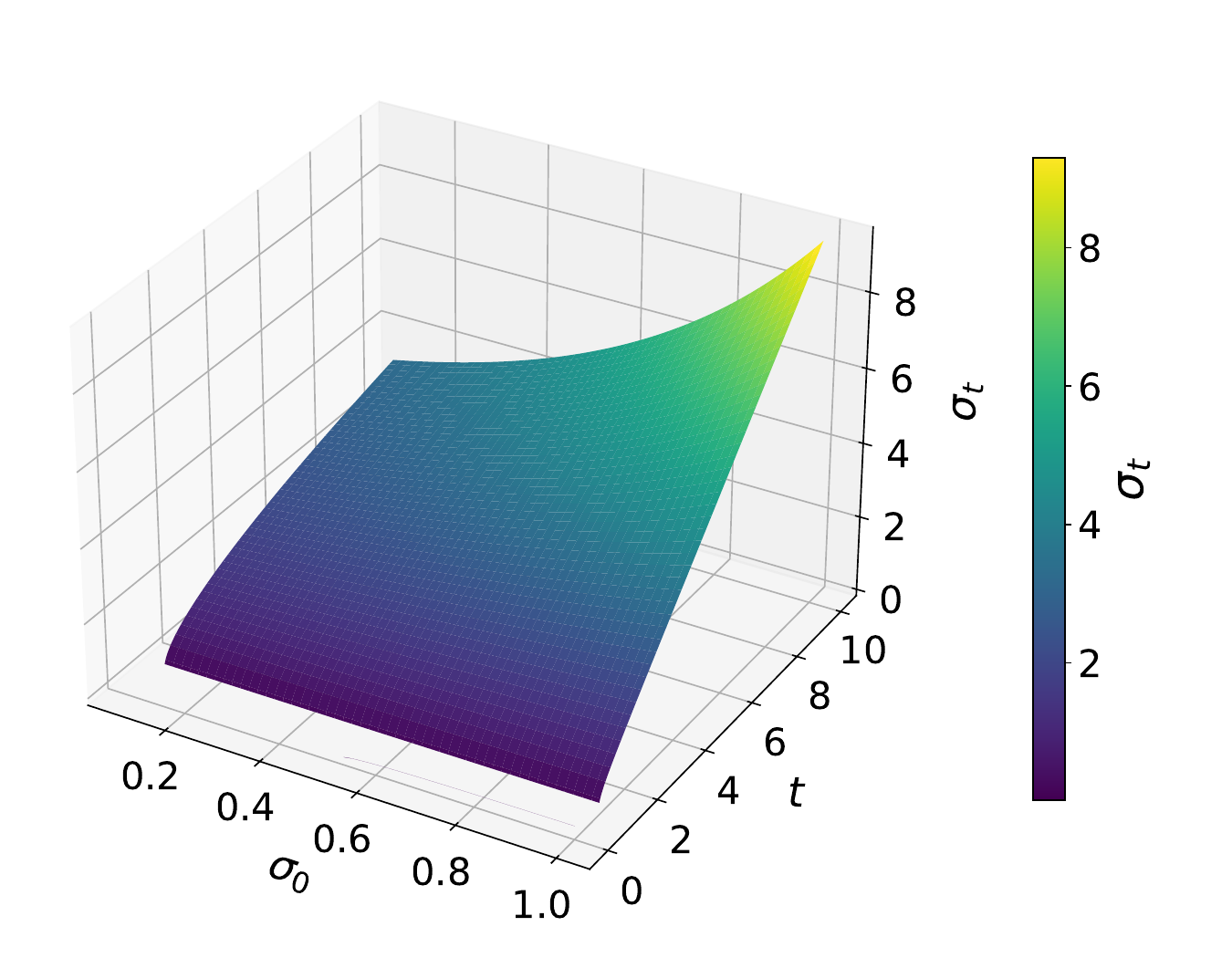}
		\caption{\textbf{Behavior of $\sigma_t$ with time $t$ and $\sigma_0$.} We plot the standard deviation $\sigma_t$ with time $t$, and $\sigma_0$ is depicted by a sequential color plot. Here the time ($t$) axes and $\sigma_0$ are taken along the horizontal plane, and $\sigma_t$ is plotted along the vertical axes. }
		\label{fig.22}
	\end{figure}
    \begin{figure*}
		\centering
	\includegraphics[scale=0.38]{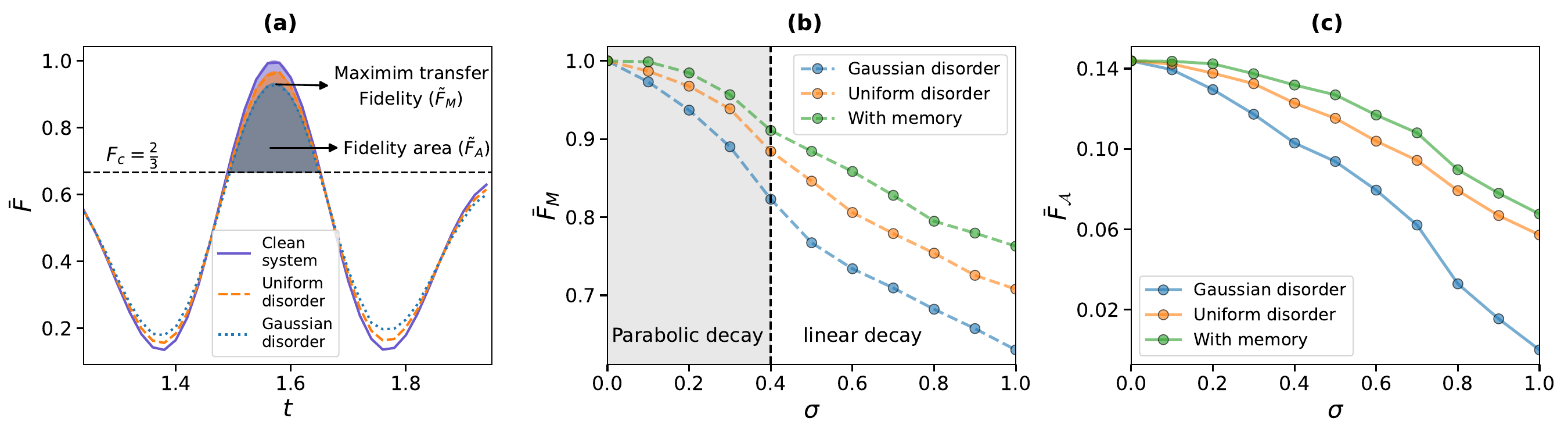}
		\caption{\textbf{Variation of maximum fidelity and fidelity area with disorder strengths.} In panel (a), we plot the variation of the transfer fidelity $\bar{F}$ with time ($t$). The curve corresponding to the clean system is represented by the smooth violet line, while the curves corresponding to uniform, Gaussian, and time-dependent memoryful disorder are shown by the dashed orange and dotted blue lines, respectively. The horizontal dashed line at $2/3$ represents the classical fidelity $F_c$. The shaded region above this line under the dynamical fidelity curve denotes the regime where the fidelity exceeds the classical fidelity.
Panel (b) shows the variation of the maximum fidelity $\bar{F}_M$ with the disorder strength $\sigma$. The curves corresponding to uniform, Gaussian, and time-dependent memoryful disorder are represented by the orange, blue, and green dashed lines with circular markers. Here the gray region left to $\sigma=0.4$ denotes the parabolic decay of transfer fidelity, and in the region right to $\sigma=0.4$, it shows linear decay. 
Panel (c) depicts the variation of {fidelity area}, $\bar{F}_A$, with $\sigma$, where fidelity area denotes the area under the dynamical fidelity curve above the classical fidelity ($F_c$) line. The curves corresponding to uniform, Gaussian, and time-dependent memoryful disorders are shown by the orange, blue, and green smooth lines with circular markers. All the curves are plotted for $\lambda=1$, $h=1/2$, and system size $L=16$.} 
		\label{fig2}
	\end{figure*}
In our work, we consider three different types of disorder: two of them are memoryless disorders, specifically uniform and Gaussian disorders, and a third class is time-dependent memoryful disorder.
Here we provide a brief discussion about memoryful disorder that evolves dynamically in time, retaining correlations with its past realizations, in contrast to conventional memoryless disorder where random parameters are independently sampled throughout the evolution from a static distribution. Motivated by the continuum limit distribution of the elephant quantum walk~\cite{elephant}, we implement memoryful disorder via time-dependent Gaussian distribution.

The time-dependent Gaussian distribution is given by
\begin{align}\label{eq4}
P(\delta,t) &= \frac{1}{\sqrt{2\pi\sigma^2_t}}
\exp\left[-\frac{(\delta-\bar{\delta})^2}{2\sigma^2_t}\right], \text{where,} \\ 
\sigma_t &= \sqrt{\sigma_0^2+2tD_s(t)}, \nonumber\hspace{0.1cm} \text{and}\hspace{0.1cm} 
D_s(t) = \frac{\left[t^{2a-1}-1\right]}{4a-2}.
\end{align}
Here $P(\delta,t)$ is the distribution at which the parameter $\delta$ is sampled at time $t$. 
On the other hand $\sigma_t$ is the standard deviation at time, $t$ where $\sigma_0$ is the standard deviation at time $t=0$. On the other hand the quantity $D_s(t)$ is called the diffusion constant. The dependence of the time-dependent standard deviation ($\sigma_t$), defined in Eq.~\eqref{eq4}, on the initial disorder strength ($\sigma_0$) and time $t$ is illustrated in Fig.~\ref{fig.22}. 



\section{Memory-assistance for quantum state transfer}\label{III}

 In this section, we examine the robustness of the quantum state transfer protocol in the presence of disorder in the many-body scar system as described in Eq.~\eqref{Eq.3}. At the same time, we compare the effects of three types of disorder, i.e., memoryless uniform, Gaussian, and memoryful disorder to determine which one is more favorable for QST.
To this end, we consider the following two figures of merit.
   

   \begin{enumerate}
  \item The first one is maximum fidelity $(\bar{F}_M)$, defined as the maximum value attained over the time domain. 
  \item 
 The second one is the fidelity area, which is the area of the region covered under the dynamical fidelity profile lying above the classical threshold value $F_c=2/3$. Several dynamical parabolic arches lie above the classical fidelity threshold, $F_c=2/3$, and for the present analysis we consider only the first arch.
 \end{enumerate}

These two figures of merit are pointed out in panel (a) of Fig.~\ref{fig2}.
The reason for putting focus on the region above the classical fidelity $F_c$ is basically that it denotes the operational region where the quantum advantage persists. This motivates us to consider the fidelity area as an additional figure of merit alongside the maximum transfer fidelity. 

   In panel (a) of Fig.~\ref{fig2}, the transfer fidelity as a function of time for memoryless uniform and Gaussian disorders along with the clean case is depicted by orange dashed, blue dotted, and violet smooth lines in order to investigate its dynamical behavior. Here we have considered the system size $L=16$. For the time dynamics, we use the Chebyshev-polynomial method~\cite{cheb_RMP,cheb_Book}. The other parameters $\lambda$, $h_z$ and $\sigma$ are chosen to be $\lambda=1$, $h_z=1/2$ and $\sigma=0.2$ respectively. The quantities $\bar{F}_M$ are obtained by averaging over $1000$ disorder realizations as well as over the initial state parameters. 
From the figure, apart from the clean system, which exhibits unit transfer fidelity, the uniform disorder fails to achieve unit fidelity but still performs better than the Gaussian disorder.

To elucidate the dependence of the state transfer protocol on the disorder strength in our considered model, we plot the maximum fidelity ($\bar{F}_M$) as a function of the disorder strength ($\sigma$) for all three types of disorder, shown in panel (b) of Fig.~\ref{fig2}. Here, the horizontal axis represents the disorder strength ($\sigma$), while the vertical axis denotes the maximum fidelity ($\bar{F}_M$). The parameter $\delta_n$, as defined in Eq.~\eqref{2eq}, is sampled from distributions with zero mean, while the disorder strength is taken to lie within the interval $\sigma \in [0,1]$. In the case of memoryful time-dependent distributions, as the disorder strength evolves with time, for the sake of a consistent and equitable comparison, we fix its initial value to be identical to that employed for the other two memoryless disorder scenarios. It is also to be noted that throughout the paper the disorder strength corresponds to the time-dependent distribution, actually implying its initial value. At the same time, the parameter $\alpha$ corresponding to the diffusion coefficient is taken as $\alpha=3/4$ throughout the paper. It is also worth emphasizing that for each time step and for every realization corresponding to both the disorder and the initial states, we numerically optimize the fidelity over the unitary introducing the local phase discussed in Sec.~\ref{IIB} and the unitary comes out to be independent of initial state. The orange and blue dashed curves with markers correspond to the memoryless uniform and Gaussian disorder cases, respectively, whereas the green curve represents the memoryful disorder case. It can be observed from Fig.~\ref{fig2} (b) that for all three types of disorders, the maximum fidelity decreases with increasing disorder strength. However, the fidelity ($\bar{F}_M$) corresponding to memoryful disorder remains consistently higher at any nonzero considered disorder strength compared to the two memoryless cases (static uniform and Gaussian disorder), thereby unequivocally establishing the enhanced performance of the QST protocol arising from the presence of memory in the disorder in comparison to memoryless cases. Moreover, it can be noted that the decrease of the maximum fidelity with increasing disorder strength shows a parabolic nature up to $\sigma \approx 0.4$ for all three types of disorder. However, for $\sigma > 0.4$, the maximum fidelity decreases almost linearly. This behavior indicates a parabolic to linear transition in the decay of maximum transfer fidelity $\sigma \approx 0.4$. 
In case of uniform disorder and time-dependent memoryful disorder, the maximum fidelity stays higher than the classical fidelity $F_c$ even at disorder strength $\sigma=1$. On the other hand, in case of Gaussian disorder, the quantum advantage persists up to the disorder strength $\sigma \approx 0.9$ roughly. 

In addition to the above analysis, which is based on Fig.~\ref{fig2} (b), we study how the fidelity area ($\bar{F}_A$), varies with the disorder strength for the three types of disorders shown in panel (c) of Fig.~\ref{fig2}. In this panel, $\bar{F}_A$ is plotted along the vertical axis, while the disorder strength $\sigma$ is shown along the horizontal axis. The range of disorder strength is chosen to be the same as in panel (b), i.e., $\sigma \in [0,1]$. A qualitatively similar trend is observed for the fidelity area $\bar{F}_A$. Among the three disorder types, memoryful disorder consistently yields the largest $\bar{F}_A$. In contrast, between the two memoryless disorder models, the uniform disorder produces higher $\bar{F}_A$ values than the Gaussian disorder.


Taken together, both figures of merit, $\bar{F}_M$ and $\bar{F}_A$, conclusively demonstrate that the incorporation of memory effects in the disorder is advantageous for QST under the dynamics of the disordered many-body scar Hamiltonian described in Eq.~\eqref{Eq.3}.

\section{Variation of degree of scarness with  disorder strength}\label{IV}

In general, for an ergodic quantum system, the dynamics of a local observable relaxes to its thermal expectation value~\cite{ETH_1,ETH_2}. However, it has been demonstrated in~\cite{scar_1} that many-body systems hosting quantum many-body scar eigenstates can exhibit periodic revivals in the dynamics of local observables.

In this section, we study the effect of disorder on the periodic revivals that appear in the dynamics of local observables in the considered many-body scar Hamiltonian and further introduce a measure of scarness based on the periodic revival behavior of local observables. Here, we consider $\sigma_z$ as the local operator corresponding to the first site of the chain. We introduce a quantity $\langle\sigma_z\rangle_t^{\text{loc}}$ which denotes the expectation value of $\sigma_z$ at time $t$, averaged over all initial states sampled Haar uniformly, such that
\begin{equation}
\langle \sigma_z \rangle_t^{\text{loc}}
=
\frac{1}{4\pi}\int \mathrm{Tr}[\rho^{\theta,\phi}_1(t)\sigma_z]\sin(\theta)\, d\theta\, d\phi.
\end{equation}
Here, $\rho^{\theta,\phi}_1(t)$ denotes the evolved state at a time $t$ corresponding to site $1$ starting from the initial state $\ket{\psi(\theta,\phi)}$. In our numerical study we use 
Monte Carlo method for the Haar uniform averaging.

\begin{figure}
		\centering
	\includegraphics[scale=0.42]{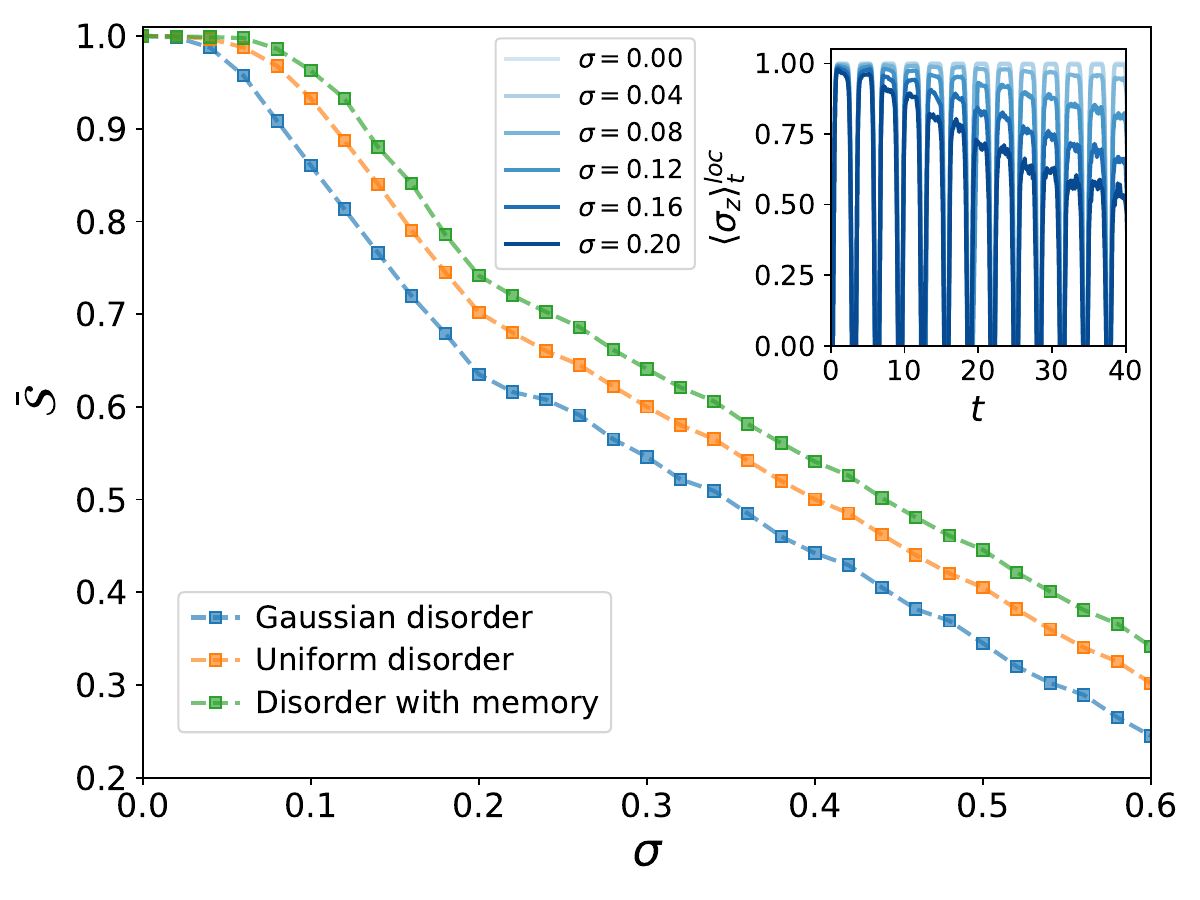}
		\caption{\textbf{Behavior of the degree of scarness as a function of disorder strength.} Here we plot $\bar{S}$ as a function of $\sigma$ for Gaussian, uniform, and memoryful disorder. The blue, orange, and green dashed curves with square markers denote Gaussian, uniform, and time-dependent (memoryful) disorder. In the inset, we plot $\langle\sigma_z\rangle^{\text{loc}}_t$ as a function of $t$ corresponding to Gaussian disorder for different disorder strengths $\sigma$. The different values $\sigma=0.00,0.04,0.08,0.12,0.16,0.20$ are represented by sequential colors with varying intensity. The whole study is done for $\lambda=1$, $h=1/2$, and system size $L=16$.}
		\label{fig:my_label}
	\end{figure}

\begin{figure*}\label{V}
		\centering
	\includegraphics[scale=0.17]{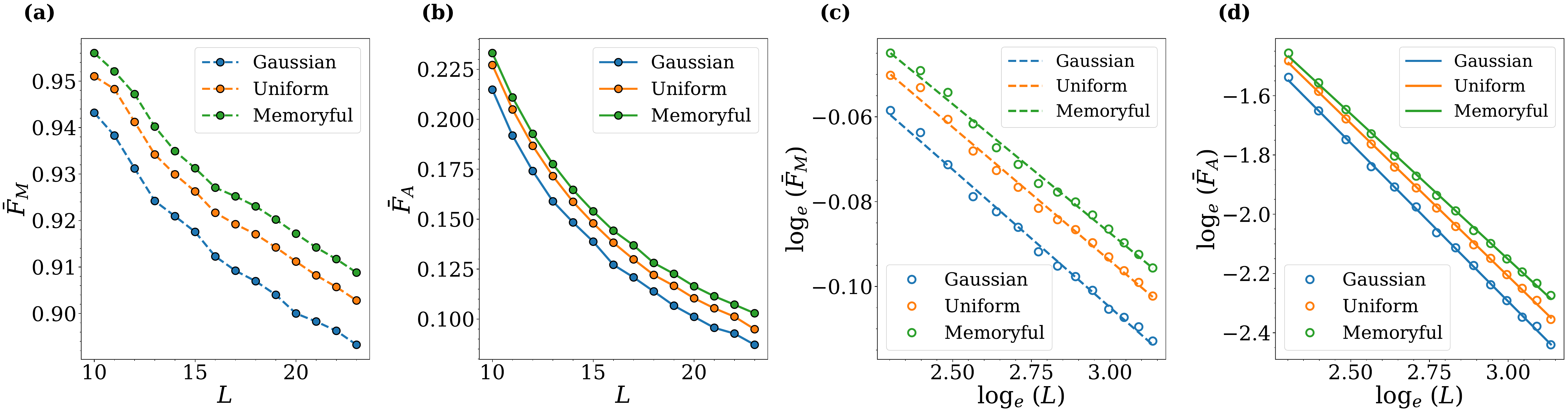}
		\caption{\textbf{Scaling of $\bar{F}_M$ and $\bar{F}_M$ with system size ($L$).} In panel (a), we present the maximum fidelity $\bar{F}_M$ as a function of the system size $L$, considering to $L=10,11,12,13,14,..,22,23$. The orange, blue, and green dashed curves with circular markers correspond to uniform, Gaussian, and time-dependent memoryful disorders, respectively. Panel (b) shows the corresponding behavior of $\bar{F}_A$ for the same range of system sizes. The data points are represented using the same type of markers and color for the three types of disorders as in panel (a), but with different line styles (smooth lines). In panel (c), we perform a scaling analysis of $\bar{F}_M$, demonstrating a clear $\log$-$\log$ scaling behavior.  A similar scaling trend is observed for $\bar{F}_A$ in panel (d). For panels (c) and (d), the raw data are denoted by circular markers, and the fitted curves are given by dashed and smooth lines, respectively. The corresponding color is provided in the legend. Throughout all panels, the disorder strength is fixed at $\sigma=0.2$ for all three disorders. The other parameters are set to $\lambda=1$ and $h=1/2$.
}
		\label{fig_4}
	\end{figure*}

In the inset of Fig.~\ref{fig:my_label}, we plot $\langle\sigma_z\rangle_t^{\text{loc}}$ vs $t$ plot, considering the system size $L=16$ and other parameters $\lambda=1$, $h=1/2$ upon introducing Gaussian disorder with disorder strengths $\sigma=0.04,0.08,0.12,0.16,0.20$ along with the clean case $\sigma=0.00$.
A prominent periodic revival is observed in the time dynamics of the local observable $\sigma_z$ for the clean case, i.e., $\sigma=0$. But we find that the periodic revivals gradually diminish as the disorder strength $\sigma$ increases. It basically implies that increasing the disorder strength, the system shows more prominent thermal behavior.  Utilizing this behavior, we introduce a measure of scarness of the disordered many-body scar Hamiltonian in terms of the peaks of the revivals. We term this measure as "degree of scarness."

Let us denote the height of the $i$'th peak in the dynamics of $\langle \sigma_z \rangle_t^{\text{loc}}$ by $l_i$, and suppose we are considering $\mathcal{N}$ numbers of peaks starting from the initial peak, where $\mathcal{N}>>1$. Then the degree of scarness is defined as
\begin{equation}
\bar{\mathcal{S}}=\frac{\sum_{i=1}^{\mathcal{N}}l_i}{\mathcal{N}}.
\end{equation}

In Fig.~\ref{fig:my_label}, we plot the $\bar{\mathcal{S}}$ vs. $\sigma$ plot for uniform, Gaussian, and memoryful disorders. The orange, blue, and green dashed curves with square markers correspond to the Uniform, Gaussian, and time-dependent disorders. For all three types of disorder, i.e., uniform, Gaussian, and memoryful disorder, we observe a non-linear to linear transition in the decay of the degree of scarness as a function of the disorder strength, occurring around the critical disorder strength $\sigma = 0.2$. The most important point is to note that the degree of scarness corresponding to memoryful disorder stays above the degree of scarness corresponding to memoryless uniform and Gaussian disorder. And in between uniform and Gaussian disorder the value of the degree of scarceness remains lower than the uniform disorder.
The degree of scarness decreases with increasing disorder strength; in other words, it indicates an enhancement of thermalization as the disorder strength increases. Furthermore, from the analysis of Fig.~\ref{fig2} (a), (b), and Fig.~\ref{fig:my_label}, we observe that a higher degree of scarceness (weaker thermalization) leads to better state-transfer performance.
Consequently, memoryful disorder exhibits superior performance compared to both uniform and Gaussian (memoryless) disorder, as it possesses a comparatively higher degree of scarness than the other two disorder types. Moreover, for the same reason, among the two memoryless disorders, uniform disorder outperforms Gaussian disorder, as it possesses a comparatively higher degree of scarness.

\section{scaling analysis}\label{SV}

In this section, we investigate the behavior of the maximum fidelity and the fidelity area as a function of the system size ($L$). To this end, in panel (a) of Fig.~\ref{fig_4}, we present the variation of $\bar{F}_M$ with $L$ for all three types of considered disorder, namely memoryful disorder, uniform disorder, and Gaussian disorder. For all three types of disorders, we consider the disorder strength $\sigma=0.2$. Here, the values of $\bar{F}_M$ in panels (a) of Fig.~\ref{fig_4}, corresponding to the uniform, Gaussian, and time-dependent memoryful disorders, are denoted by orange, blue, and green markers with dashed lines with circular markers, respectively. The system sizes are considered $L=10,11,12,13,14,...,22,23$. For all three types of disorder, the value of $\bar{F}_M$ decreases with the system size $L$ non-linearly. We observe that, in the presence of time-dependent memoryful disorder, the state transfer protocol exhibits superior performance compared to both uniform and Gaussian disorder by possessing a higher value of $\bar{F}_M$ across all considered system sizes. Furthermore, between the two memoryless disorder profiles, uniform disorder consistently outperforms Gaussian disorder for all the considered system sizes.

At the same time, in panel (b) of Fig.~\ref{fig_4}, we plot fidelity area ($\bar{F}_A$) as a function of system size ($L$) for memoryful, uniform, and Gaussian disorder, considering the same system sizes as in panel (a). The data corresponding to memoryful, uniform, and Gaussian disorder are represented by orange, blue, and green smooth line circular markers, respectively. We observe the same qualitative behavior in $\bar{F}_A$ as in $\bar{F}_M$, namely that memoryful disorder yields higest value of $\bar{F}_A$, followed by uniform disorder and then Gaussian disorder.

To analyze the decay behavior of $\bar{F}_M$ with respect to the system size $L$ more clearly, we perform curve fitting using an $\log$-$\log$ scaling function for all three types of disorder, is given by
\begin{equation}
   \log_e(y)=m\log_e(x)+c\nonumber .
\end{equation}
Here, $x$ and $y$ correspond to the quantity system size ($L$) and maximum transfer fidelity ($\bar{F}_M$), respectively. The fitted curves corresponding to memoryful, uniform, and Gaussian disorder are denoted by green, orange and blue dashed lines in panel (c) of Fig.~\ref{fig_4}. 
Similarly, we analyze the scaling of $\bar{F}_A$ with the system size $L$ by fitting the data points in panel (d) of Fig.~\ref{fig_4} to the logarithmic relation $\log(y') = m'\log(x') + c'$. The quantity $x'$ and $y'$ denotes the fidelity area $\bar{F}_A$ and system size $L$. The fitted curves corresponding to memoryful, uniform, and Gaussian disorder are denoted by Green, orange, and blue smooth lines. The fitted parameters $\{m,c\}$ and $\{m',c'\}$ and the least squares error (LS) corresponding to panels (c) and (d) for all three types of disorders are shown in table~\ref{table_1}.

\begin{table}[h]
\centering
\renewcommand{\arraystretch}{1.25}

\begin{tabular}{c|ccc|ccc}
\hline
 & \multicolumn{3}{c|}{$\mathbf{\bar{F}_M}$} 
 & \multicolumn{3}{c}{$\mathbf{\bar{F}_A}$} \\
\hline
\textbf{Disorder} & $c$ & $m$ & $LS(\%)$ & $c'$ & $m'$ & $LS(\%)$ \\
\hline
Gaussian 
& $0.0902$ & $-0.065$  & $0.124$ 
& $0.886$ & $-1.068$ & $0.496$ \\
\hline
Uniform  
& $0.0941$ & $-0.062$  & $0.11$  
& $0.901$ & $-1.031$ & $0.715$ \\
\hline

\makecell[c]{Memoryful\\ disorder} 
& $0.0950$ & $-0.0608$ & $0.118$ 
& $0.912$ & $-0.981$ & $0.675$ \\
\hline

\end{tabular}

\caption{Fitted parameters and corresponding least-squares errors for different disorder profiles.}
\label{table_1}
\end{table}

From the table, it is evident that the slopes $m$ and $m'$ corresponding to the memoryful disorder have the smallest magnitudes, while the constants $c$ and $c'$ attain the largest values compared to those associated with the other disorder profiles. It implies the beneficial role of the memory effect in disorder in enhancing the Proficiency of quantum state transfer even in larger distances in one-dimensional disordered spin chains hosting scar eigenstates.

\section{Role of inverse partition ratio in quantum state transfer}\label{VI}

To obtain a more clearer understanding of the relation between ergodicity and the state transfer protocol, we employ a popular measure of ergodicity, namely the inverse participation ratio (IPR).

Let us consider a normalized many-body quantum state expanded in the basis $\{\ket{i}\}$ as
$
\ket{\psi} = \sum_{i=1}^{D} c_i \ket{i},
$
where $D$ denotes the dimension of the corresponding Hilbert space $\mathcal{H}_D$ and $c_i$ are the amplitudes. The inverse participation ratio (IPR) is then defined as

\begin{equation}
\text{IPR} = \sum_{i=1}^{D} |c_i|^4 .
\end{equation}

For nonergodic or integrable systems, the value of IPR remains finite, but for ergodic systems, it is almost equal to $\text{IPR}\approx1/D$~\cite{IPR_algor}.

\begin{figure}
		\centering
	\includegraphics[scale=0.24]{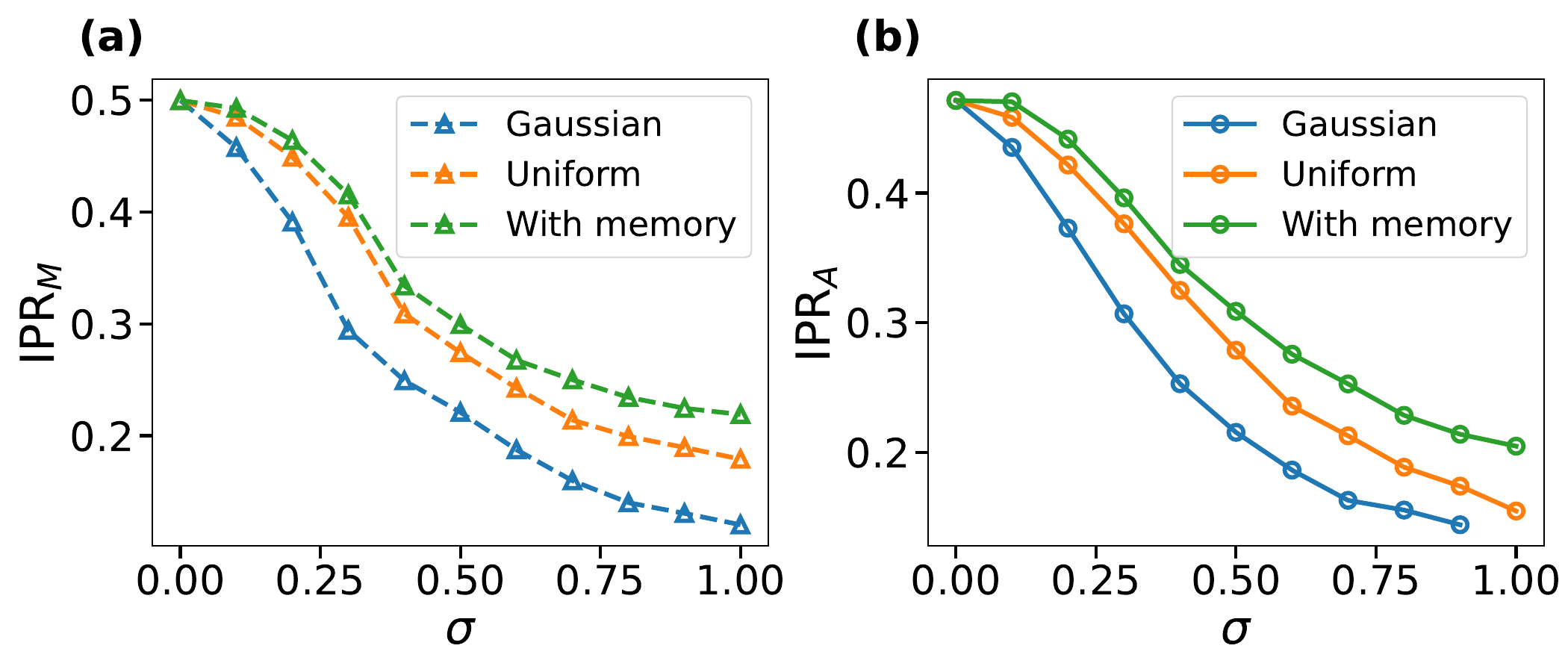}
		\caption{\textbf{Behavior of inverse participation ratio with disorder strength.} In panel (a), we plot $\text{IPR}_M$ as a function of the disorder strength $\sigma$. The results corresponding to uniform disorder, Gaussian disorder, and memoryful disorder are represented by orange, blue, and green dashed curves with triangular markers, respectively. 
In panel (b), we present $\text{IPR}_A$ as a function of $\sigma$ for the same three types of disorder, denoted by red, blue, and green smooth curves with circular markers. Here, $\text{IPR}_A$ represents the average inverse participation ratio computed over the set of states whose fidelity exceeds the classical threshold $F_C$. 
In all numerical simulations, we consider a system size $L=16$, with parameters $\lambda=1$ and $h=1/2$.}
		\label{fig.5}
	\end{figure}
In this section, we examine how the inverse participation ratio (IPR), evaluated at time $t=\pi/\lambda$ (the instant at which the fidelity attains its maximum value), varies with the disorder strength $\sigma$ for all three types of disorders considered in this work. The IPR at $t=\pi/\lambda$ is denoted by $\mathrm{IPR}_M$. In panel (a) of Fig.~\ref{fig.5}, we present the variation of $\mathrm{IPR}_M$ with $\sigma$ for memoryless uniform, Gaussian, and time-dependent memoryful Gaussian disorder, represented by orange, blue, and green dashed lines with triangular markers, respectively. It is observed that, for the whole considered range of disorder strength $\sigma$, the value of $\mathrm{IPR}_M$ in the presence of memoryful disorder exceeds that obtained for both memoryless uniform and Gaussian disorders. Furthermore, among the two memoryless cases, the Gaussian disorder yields a lower $\text{IPR}_M$ compared to the uniform disorder for all considered ranges of disorder strength.

This indicates that the system exhibits stronger thermalization in the presence of Gaussian disorder compared to both uniform disorder and memoryful disorder. Moreover, when comparing uniform disorder with time-dependent memoryful disorder, the tendency toward thermalization remains more pronounced in the case of uniform disorder. As a result, the state transfer protocol exhibits greater robustness in the presence of time-dependent memoryful disorder, followed by uniform disorder, and is least robust under Gaussian disorder. Moreover, the systematic decrease of the $\text{IPR}_M$ with increasing disorder strength across all three disorder types further substantiates this observation, indicating that the ergodic behavior becomes more pronounced with the disorder strength, leading to stronger thermalization of the system and, consequently, a deterioration of the state-transfer performance.

On the other hand, we also compute the average IPR of the states whose fidelity exceeds the classical fidelity $F_C$ and is denoted by $\text{IPR}_A$. The aim of this study is to characterize the behavior of the average IPR of those states that exhibit a quantum advantage in the state transfer protocol. In panel (b) of Fig.~\ref{fig.5}, we plot $\text{IPR}_A$ as a function of $\sigma$, where the values corresponding to memoryful disorder, memoryless uniform, and Gaussian disorders are represented by green, orange, and blue smooth curves with circular markers, respectively. 
Similar to $\text{IPR}_M$, $\text{IPR}_A$ exhibits the same qualitative behavior; in particular, for a fixed disorder strength, the value of $\text{IPR}_A$ corresponding to memoryful disorder remains higher than that for memoryless disorders, and furthermore, among the two memoryless cases, uniform disorder yields a higher $\text{IPR}_A$ compared to Gaussian disorder.
\section{Conclusion}\label{VII}

Quantum many-body scars have been shown to provide a significant advantage in QST protocols by enabling the transfer of an arbitrary quantum state with unit fidelity across arbitrary distances in clean system~\cite{PST_scar}.

In this work we consider a more realistic scenario to observe the state transfer protocol in the presence of disorder by considering three different types of disorders: one is a time-dependent memoryful disorder, and the other two are memoryless disorders, i.e., uniform and Gaussian disorders, and also investigate how memory effects in disorder can provide an advantage in the QST protocol. Here, we consider two different figures of merit: the first is the maximum fidelity, defined as the maximum value of the transfer fidelity over the entire time evolution, while the second is the fidelity area, defined as the area of the dynamical fidelity curve lying above the classical fidelity threshold. We show that, although both figures of merit decrease with increasing disorder strength for all three types of considered disorders, the presence of memory in the disorder provides a genuine advantage in the state transfer protocol compared to the other two types of disorders, as captured by both figures of merit. Additionally, for all three types of disorder, the fidelity initially decreases nonlinearly up to a certain disorder strength, beyond which it decays linearly, resulting in a nonlinear-to-linear transition of the decay of transfer fidelity with increasing disorder strength. Furthermore, to explore the behavior of scarness of the system  with disorder strength and to investigate its role in (QST), we introduce a quantitative measure of degree of scarness. We observe that, for all three types of disorder, the degree of scarness remains nearly unchanged up to a certain disorder strength; beyond that, it starts to decrease. Moreover, the degree of scarness corresponding to the memoryful disorder remains higher than that of the two memoryless disorders, while between the memoryless cases, uniform disorder possesses a higher value than Gaussian disorder. This clearly indicates that a higher degree of scarness leads to better state transfer performance. After that, we perform a scaling analysis of both quantities with respect to system size and observe a clear $\log$-$\log$ scaling behavior for both the maximum transfer fidelity and the fidelity area, also showing that both figures of merit decrease with system size. Further smaller value of the slope of the fitted curve corresponding to a memoryful disorder in comparison to the other disorders clearly indicates the advantage of QST in a larger distance. To gain deeper insight into the relation between ergodicity and state transfer, we study the behavior of the inverse participation ratio corresponding to the state with maximum transfer fidelity ($\text{IPR}_M$) and the average inverse participation ratio of states with fidelity above the classical threshold ($\text{IPR}_A$) as functions of disorder strength. We find that both $\text{IPR}_M$ and $\text{IPR}_A$ decrease with increasing disorder strength for all three types of disorder, indicating enhanced ergodic behavior and a consequent degradation of the state transfer performance. At the same time, the comparatively higher values of $\text{IPR}_M$ and $\text{IPR}_A$ corresponding to the memoryful disorder, in comparison to the memoryless ones, provide a clear justification for its advantage in the state transfer protocol.



\bibliography{ref}{}
\end{document}